\documentclass[notitlepage,prl,twocolumn,nofootinbib]{revtex4-1}

\usepackage{graphicx}
\usepackage{amsmath}
\usepackage{amssymb}
\usepackage{ae} 
\usepackage{aecompl}
\usepackage{bm} 
\usepackage[utf8]{inputenc}
\usepackage[usenames,dvipsnames]{color}

\newcommand{\dm}[1]{{\Delta m^2_{#1}}}
\def\be#1{\begin{equation}\label{#1}}
\newcommand{\ee}{\end{equation}}

\begin{document}

\title{The benefits of a near detector for JUNO}
\begin{abstract}
It has been proposed to determine the mass hierarchy of neutrinos by
exploiting the beat between the oscillation frequencies corresponding
to the two neutrino mass squared differences. JUNO is based on this
concept and uses a large liquid scintillator detector at a distance of
53\,km from a powerful nuclear reactor complex. We argue that the
micro-structure present in antineutrino fluxes from nuclear reactors
makes it essential to experimentally determine a reference spectrum
with an energy resolution very similar to the one of JUNO.
\end{abstract}

\author{David V. Forero~$^{1,2}$}
\email{dvanegas@ifi.unicamp.br}
\author{Rebekah Hawkins~$^2$}
\email{rebhawk8@vt.edu}
\author{Patrick Huber~$^2$}
\email{pahuber@vt.edu}

\affiliation{$^1$~Instituto de F\'{i}sica Gleb Wataghin - UNICAMP,
13083-859, Campinas, SP, Brazil}
\affiliation{$^2$~Center of Neutrino Physics, Virginia Tech, Blacksburg, USA}

\date{\today}
\maketitle

The Jiangmen Underground Neutrino Observatory (JUNO) comprises a 20\,kt
liquid scintillator detector with a broad physics
program~\cite{An:2015jdp}. One of the key physics goals of JUNO is the
determination of the so-called mass hierarchy of neutrinos, that is, whether the third mass eigenstate is the lightest (inverted
hierarchy, IH) or heaviest (normal hierarchy, NH) without having to
rely on matter effects: with an appropriate choice of experimental
parameters, it is possible to \emph{simultaneously} be sensitive to
oscillations corresponding to the two mass squared differences
$\dm{31}\simeq 2.5\times10^{-3}\,\mathrm{eV}^2$ and $\dm{21}\simeq
7.4\times10^{-5}\,\mathrm{eV}^2$~\cite{Olive:2016xmw}. In this case,
also, the beat frequency between the two oscillations will be present,
and whether this beat frequency is larger or smaller than the main
oscillation driven by $\dm{31}$ is a direct measure of the mass
hierarchy~\cite{Petcov:2001sy}; the amplitude of the beat is given by
$\sin^22\theta_{13}$. The relative difference between the beat
frequency and main oscillation is of the order $\dm{21}/\dm{31}\simeq
1/30$, and thus an energy resolution of approximately $3\%$ is
required.

JUNO will detect reactor antineutrinos via inverse beta decay (IBD)
and is situated 53\,km from both the Yangjiang and Taishan nuclear
power plants, the distance being carefully chosen to fulfill above
conditions and to optimize the mass hierarchy sensitivity. Neutrino
energy reconstruction in IBD, $\bar\nu_e+p\rightarrow n +e^+$, is
relatively straightforward; the visible energy of the positron,
$E_{e^+}$, is related to the neutrino energy $E_\nu$ via
\be{eq:reco}
E_\nu=E_{e^+}+(m_n-m_p-m_e)\,,
\ee
neglecting the kinetic energy of the outgoing neutron, which, for the
neutrino energies in question, is an excellent approximation.  The
energy deposited by the positron in the scintillator is converted to
light, and the energy resolution is determined by photon counting
statistics and, to first order, scales as $1/\sqrt{E_{e^+}}$. 

Attaining the requisite $3\%$ energy resolution is a challenge in
itself, but, given the relatively small size of the effect, the question
of systematic uncertainties needs to be addressed. There are three main
types of systematics for this measurement: uncertainty in $\dm{31}$,
{\it i.e.} uncertainty about the main oscillation frequency;
uncertainty in the detector energy response, {\it e.g.} a shift of the
energy scale has the the same effect as a change in $\dm{31}$; reactor
antineutrino flux uncertainties, since one is searching for a
high-frequency component in the Fourier spectrum, and any
high-frequency components in the flux can lead to confusion. All of
these have been discussed in various combinations in the literature,
and, in particular, the JUNO collaboration is clearly aware of
them~\cite{An:2015jdp}.

The goal of this letter is to highlight the impact of one source of
uncertainty which is recognized but may have been underestimated: the
reactor antineutrino flux. For all the systematic effects except the
$\dm{31}$ uncertainty, some type of parameterization or implementation
in the analysis has to be found.  We argue, here, that the
parameterizations used previously to account for the reactor flux
uncertainties are not capturing the relevant degrees of uncertainty;
in other words, some of the known unknowns are not accounted for.
Using a more physical model for the reactor flux uncertainties, a
large reduction of sensitivity to the mass hierarchy results. This
problem can be completely resolved by using a reference reactor
antineutrino spectrum measured with \emph{a similar energy resolution}
as the JUNO detector. We demonstrate this by including a near detector
in the simulation.

Reactor antineutrino fluxes have taken center stage as a research
subject of their own since a series of
papers~\cite{Mueller:2011nm,Huber:2011wv} in 2011 (Huber+Mueller
model), which led to a revision of the flux models, giving rise to the
reactor antineutrino anomaly
(RAA)~\cite{Mention:2011rk}. Antineutrinos from reactors are not made
directly in the fission process; instead, they arise from the beta
decays of neutron-rich fission products.  There are roughly $10^3$
isotopes with about $10^4$ individual beta decay branches which would
have to be known with good accuracy to compute the antineutrino flux
with percent level errors. This knowledge does not exist, and thus
measurements of the total beta decay spectrum from fission
fragments~\cite{VonFeilitzsch:1982jw,Schreckenbach:1985ep,Hahn:1989zr}
are used as a basis for unfolding the antineutrino spectrum. The 2011
papers have triggered significant follow-up work, and this led to the
understanding that first forbidden non-unique beta decays, which make
up anywhere between 20-30\% of all antineutrinos relevant for IBD, and
their higher-order corrections, like weak magnetism, are dominating
the error budget and likely exceed the
estimates~\cite{Mueller:2011nm,Huber:2011wv}, a point driven home by
the observation of a 5\,MeV bump in the measured antineutrino spectrum
relative to predictions, see for instance the Daya Bay
result~\cite{An:2016srz}; for recent reviews on this
topic, see Refs.~\cite{Huber:2016fkt,Hayes:2016qnu}.

\begin{figure}
\includegraphics[width=\columnwidth]{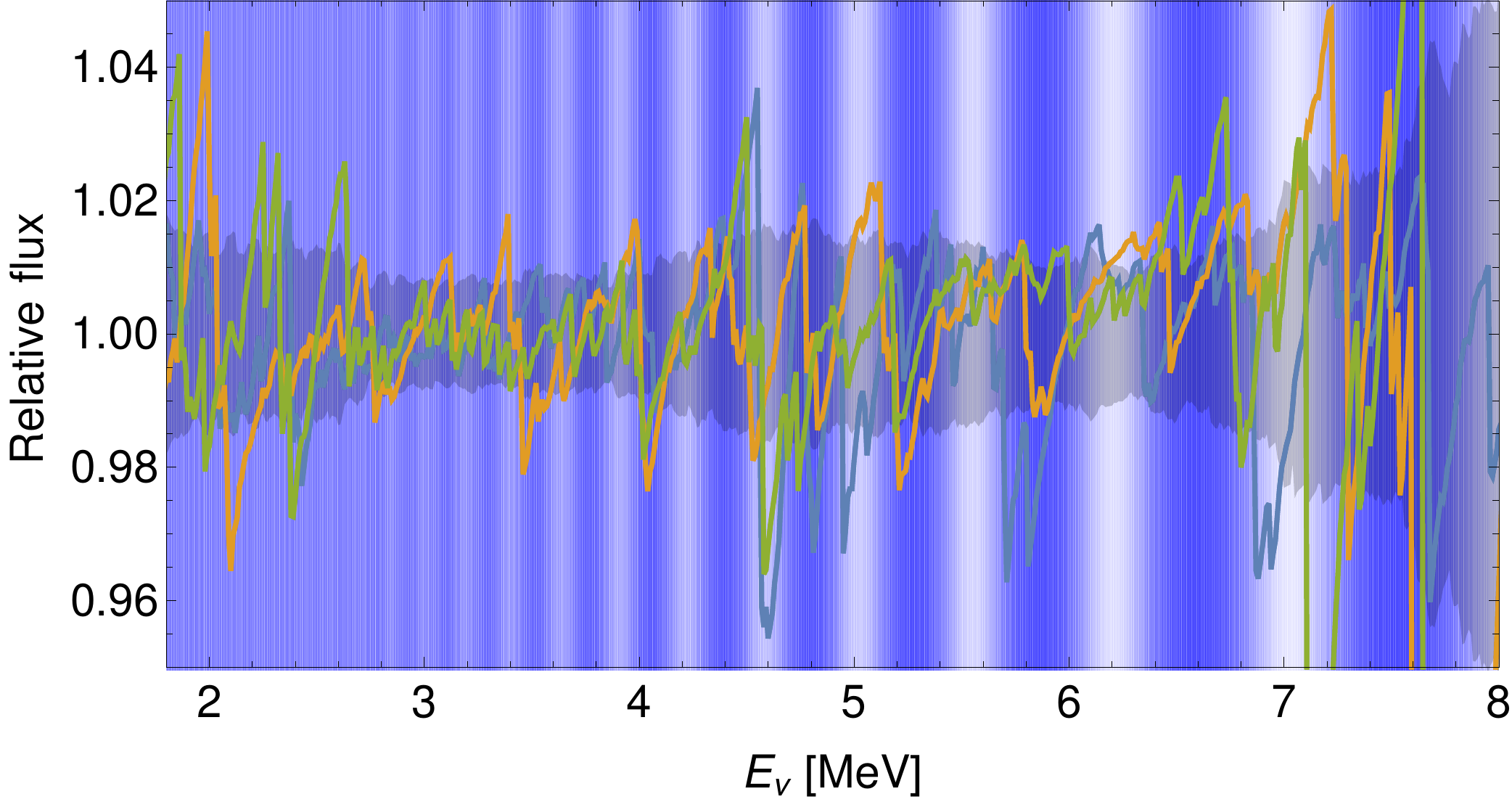}
\caption{\label{fig:shape} Shown are three synthetic antineutrino
  spectra relative to the antineutrino spectrum predicted from the ILL
  data~\cite{Mueller:2011nm,Huber:2011wv} (Huber+Mueller model)
  normalized to the same total IBD event rate. The gray shaded
  horizontal band shows the standard deviation of the whole population
  of synthetic spectra.  For illustration, the vertical bands indicate
  the oscillation arising from $\dm{31}$ at a distance $L=53$\,km
  smeared with an energy resolution of $3\%/\sqrt{E}$. }
\end{figure}

As explained, a direct calculation of antineutrino fluxes is not
feasible; nonetheless, these direct or {\it a priori} calculations
allow some significant insight into the energy structure of the
antineutrino spectrum without being obstructed by real-world detector
effects, see for instance Ref.~\cite{Fallot:2012jv}.  In Fig.~3 of
Ref.~\cite{Dwyer:2014eka}, it is highlighted that there is significant
micro-structure in the antineutrino spectrum at the 50--100\,keV
scale. This sawtooth shape arises because, in a single beta decay,
there is a finite probability to emit an antineutrino with an energy
corresponding to the entire available $Q^2$ of the transition due to
the Coulomb correction experienced by the outgoing electron. Adding a
large number of these, then, results in the sawtooth pattern also
visible in Fig.~\ref{fig:shape}.  In Ref.~\cite{Dwyer:2014eka}, it is
also shown that, once this sawtooth spectrum is convoluted with a
detector energy resolution typical for current reactor neutrino
experiments, an entirely smooth spectrum results.

{\it A priori} calculations account for about 80-90\% of all beta
decays and, thus, reproduce the total beta spectrum as measured by the
ILL experiments to about the same
degree~\cite{Fallot:2012jv}. Therefore, there is no reason to expect
that the specific energy micro-structure derived from any of these
calculations is the actual one: the specific location and size of each
sawtooth is likely wrong; the distribution of locations and sizes, on
the other hand, will be close to the true one. For the following, we use
a model based on thermal neutron fission yields of $^{235}$U,
$^{239}$Pu, and $^{241}$Pu from the JEFF database,
version~3.1.1~\cite{jeff}, and the fast neutron fission yield of
$^{238}$U from ENDF-349~\cite{lanl}. We use the beta decay information
contained in the Evaluated Nuclear Structure Data File (ENSDF)
database, version~VI~\cite{ensdf}, and the neutrino spectrum is
computed following the prescription in Ref.~\cite{Huber:2011wv}. We
use this information on fission yields and beta decays to construct a
probability density function $p(Q,a)$ for the $Q$-value and amplitude
$a$ for each beta decay branch. We then draw at random pairs of values
for $Q,a$ and compute the resulting antineutrino spectrum; we stop
adding more pairs as soon as we have the same number of antineutrinos
above IBD threshold as in the Huber+Mueller model. The resulting
antineutrino spectrum is then normalized to the same IBD rate as
obtained from the Huber+Mueller model and reweighted to represent the
shape of the Huber+Mueller flux at an energy resolution of
$8\%/\sqrt{E}$. We repeat the procedure 1000 times to obtain a
population of synthetic antineutrino spectra which all correspond to a
very similar spectrum at $8\%/\sqrt{E}$ resolution. The results are
shown in Fig.~\ref{fig:shape}, where we show the resulting
distribution for each energy bin relative to its mean (the
Huber+Mueller prediction), and we also show three examples of a
synthetic spectrum. For comparison, the relevant oscillation is
overlayed; it is apparent that much of the structure in the synthetic
spectrum is at a similar frequency and amplitude as the effect sought
after in JUNO. This indicates that the strategy outlined in
Ref.~\cite{An:2015jdp} to deal with the reactor flux uncertainty,
namely to use the Daya Bay measured spectrum as reference spectrum is
fraught with difficulty: the Daya Bay spectrum has been measured with
an energy resolution of approximately $8\%/\sqrt{E}$ whereas for JUNO
the spectrum at $3\%/\sqrt{E}$ is needed.

The question now is: what energy resolution does the reference
spectrum need to be measured with, and what other detector effects
could intervene? The effect of having a second detector (near
detector) in the hierarchy determination has been discussed in more
general terms in
Refs.~\cite{Ciuffoli:2013pla,Capozzi:2015bpa,Wang:2016vua}. Specifically,
we investigate the non-linearity of the energy response as a potential
issue in comparing the data from two detectors. The Daya Bay detectors
are precision instruments, and their success has inspired the design
for JUNO, therefore it makes sense to use them as a proxy for the
energy response. For the Daya Bay detectors, the
energy response to positrons is non-linear with the main effect
happening below the $4\,\text{MeV}$.  This effect is attributed to
ionization quenching of scintillation light and Cerenkov light
production and peculiarities of the electronics~\cite{An:2013zwz}.

To include non-linear effects in the reconstruction of the positron
energy one can parameterize the effect as a linear combination of
functions that are powers of energy, generalizing the linear
scaling~\cite{Li:2013zyd,Capozzi:2015bpa}:
\be{eq:energySeries}
\frac{E_{\text{rec}}}{E}=1+\sum_{k=0}^n \alpha_k E^k\equiv 
1+\delta_{\text{scal}}(E).
\ee
For $k=0$, and with only $\alpha_0\ne0$, one obtains the linear case
(linear scaling). Therefore, to include non-linear effects,
$\alpha_k\ne0$, one needs to include higher energy powers in
Eq.~(\ref{eq:energySeries}). Even though this approach is conceptually
simple, it does offer neither a physical condition when to stop the
series nor allows to set the size and to understand the physical
meaning of the non-linear coefficients.

Assuming the energy response of the JUNO detector will be similar to
the one of Daya Bay Ref.~\cite{An:2016ses}, it is possible to estimate
the remaining error $S_{\text{scal}}(E)$ after the non-linear
correction has been applied as a relative deviation from the nominal
model. This procedure was implemented in
Ref.~\cite{Li:2013zyd,Capozzi:2015bpa}, where the error envelope
$S_{\text{scal}}(E)$ is taken from the reported detector energy
response function in Ref.~\cite{An:2016ses}. The largest energy scale
error is around $2\%$ in the low energy region; for most of the energy
range the error is below $1\%$; we follow the procedure outlined in
Ref.~\cite{Capozzi:2015bpa}.

To include the effect of the remaining error after the non-linear
detector energy response correction in the analysis, we have
generalized the algorithm used to account for the linear scaling in
GLoBES ~\cite{Huber:2003pm}. Notice that in principle the individual
$\alpha_k$ coefficients can be arbitrarily large, except for
$|\alpha_0|\lesssim 2\%$, since they have no direct physical meaning,
and, at any energy, only the sum of all terms is constrained by the Daya
Bay model. This is clearly a shortcoming of the parameterization given
in Eq.~(\ref{eq:energySeries}). The overall Daya Bay constraint is
implemented as a penalty on the $\chi^2$-function~\cite{Capozzi:2015bpa}:
\be{eq:penalty}
\chi^2_{\text{scal}}=\text{max}_E\left|\frac{\delta_{\text{scal}}(E)}{S_{\text{
scal}}(E)}\right|^2.
\ee
After implementing the penalty in Eq.~(\ref{eq:penalty}) in the JUNO
simulation, we find that the $\chi^2$-function becomes somewhat wider
as a function of $\dm{31}$, indicating as mentioned previously, the
similarity between a linear energy scale uncertainty and a less
precise input on $\dm{31}$. We also find that the result practically
does not change for $k=2$, and thus there is no reason to go beyond
$k=2$. For $k=2$ the sensitivity to the mass hierarchy is decreased by
$\sim 1.23$ $\Delta \chi^2$ units, taking the case of the linear
scaling as reference ($\Delta \chi^2_{\text{min}}\simeq23$), which is
compatible with the result in Ref.~\cite{Capozzi:2015bpa}. Also, a
pull on $\Delta m_{31}^2$ has to be included~\cite{Li:2013zyd}, since
its effect (within the current errors) can mimic the non-linear
effects we are introducing, as was pointed out
previously~\cite{Qian:2012xh,Ciuffoli:2013pla}. 

\begin{figure}
\includegraphics[width=\columnwidth]{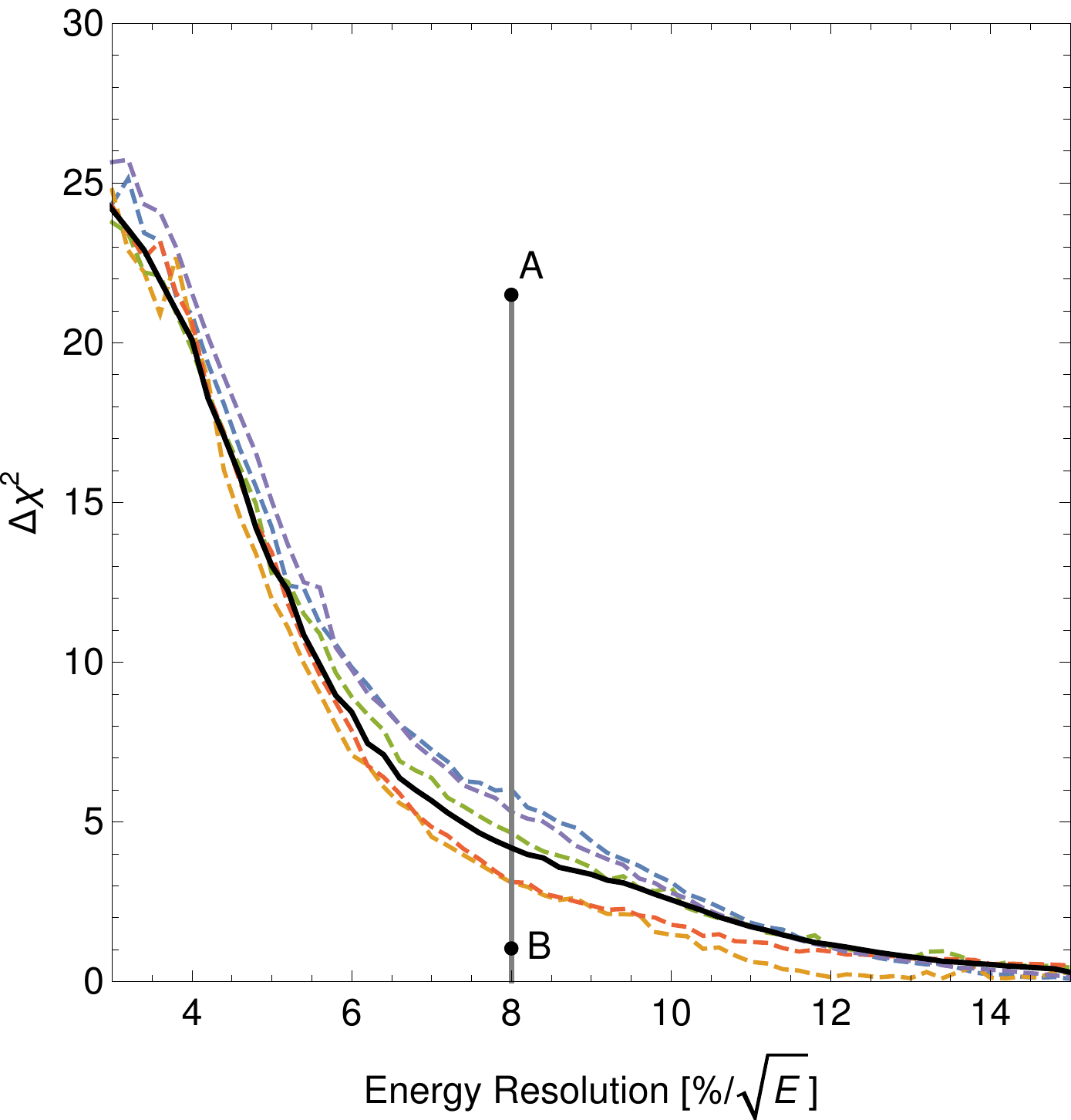}
\caption{\label{fig:result} $\Delta \chi^2$ between normal and
  inverted hierarchy (assuming normal hierarchy to be true) for a
  JUNO-like experiment employing a reference measurement of the
  reactor spectrum with an energy resolution given by the
  abscissa. The solid line assumes that the same spectrum is used to
  generate and fit the data, whereas, for the dashed lines, we use a
  random synthetic spectrum to generate the data and fit the data with
  Huber+Mueller model. The points labeled A and B, show the result
  obtained by using the actually measured antineutrino spectrum from
  Daya Bay~\cite{An:2016srz}: for point A we are using one nuisance
  parameter per Daya Bay bin and for point B we are using one nuisance
  parameter in each of 100 bins.}
\end{figure}

Now, we can address how well a reference spectrum needs to be measured
in terms of energy resolution. To this end, we set up a model with a
5\,ton near detector at a distance of 0.5\,km, which yields
approximately $1.6\times10^5$ events per year. For the far detector,
we assume 20\,kt fiducial mass at a distance of 53\,km. For both
detectors, we use the non-linearity model described above with $k=2$,
and the $\alpha_k$ are varied independently for near and far
detectors, reflecting the assumption that the $\alpha$'s are
constrained by the calibration systems of each detector
independently. We use 100 bins to compute the spectrum prior to
applying the energy resolution function, and for each bin, we
introduce a nuisance parameter, which is fully correlated between near
and far detectors, but otherwise unconstrained. This corresponds to a
flux model where no prior knowledge on fluxes is assumed except that
the energy scale of variations can be as small as about 50\,keV. We
argued in the introduction that the relevant degrees of uncertainty in
the flux lie in this micro-structure, and, locally, deviations can be
quite large, see Fig.~\ref{fig:shape}, whereas the deviation of the
mean is quite small. Finally, we use $3\%/\sqrt{E}$ as energy
resolution for the far detector and vary the resolution of the near
detector, with the result shown as solid line in
Fig.~\ref{fig:result}. For the solid line, we use the same flux model
(Huber+Mueller) to generate the data and fit it. For the dashed lines,
we use a synthetic spectrum, drawn at random from the population, to
generate the data and attempt to fit it with the Huber+Mueller model:
the result is close to the solid line. We conclude that, without a
dedicated near detector, the sensivity accounting for realistic flux
errors decrease from $\Delta\chi^2\simeq 22$ to $\Delta\chi^2\simeq4$;
whereas, with a decicated near detector, the sensitivity may improve
beyond the original one, since all flux uncertainties are eliminated.

For comparsion, in Fig.~\ref{fig:result} we also show the
$\chi^2$-values obtained by using instead of a near detector the
actually measured Daya Bay antineutrino spectrum and its full
covariance matrix~\cite{An:2016srz}. Point A corresponds to the case
where we allow one nuisance parameter for each Daya Bay energy bin of
approximately 250\,keV width and find indeed, that the Daya Bay data
would fully eliminate the effect of flux uncertainties. Point B,
however, is computed with one nuisance paramater for each of 100 bins
corresponding to the case of significant micro-structure in the
antineutrino flux. Clearly, in that case the Daya Bay measurement is
\emph{insufficient} to eliminate this systematic. Note, that point B
lies below the black line, which is based on a idealized near
detector, because of real-world detector effects which result in
significant spectrum uncertainties in the Daya Bay data for very low
and very high antineutrino energies.

In summary, we made the argument that antineutrino reactor fluxes have
a micro-structure at the 50\,keV level which is similar to the mass
hierarchy signal in experiments like JUNO. We performed a careful
study of the impact of non-linearities in the energy response and find
that, overall, they  have a limited impact on these experiments. We show
that a reference measurement of the antineutrino spectrum with an
energy resolution very similar to the far detector is needed to
exclude any sensitivity reduction due to the unknown micro-structure of
the antineutrino flux. For simplicity, we implement a near detector,
which, for a real experiment, has the added benefit of providing
information on the actual running conditions of the reactor(s), and
thus helps to eliminate any systematic uncertainties which otherwise
could arise. However, a reference measurement performed at a different
time and/or different reactor, for the purposes of this study, would
be acceptable as long as the energy resolution is similar to the one of
the far detector.


\acknowledgments

This work was supported by the U.S. Department of Energy under awards
\protect{DE-SC0009973} and \protect{DE-SC0018327}. DVF is thankful for
the support of S\~ao Paulo Research Foundation (FAPESP) funding Grant
No. 2014/19164-6 and 2017/01749-6., and FAEPEX found agency No 2391/17
for partial support.

\bibliography{./references.bib,./mass_ordering_JUNO}

\end{document}